\newif\ifuseprd
\newif\ifom
\useprdtrue 
\omfalse    
\ifuseprd 
\documentclass[preprint,tightenlines,floats,prd,eqsecnum,nobibnotes,%
nofootinbib]{revtex4}
\else
\documentclass[final,12pt,letterpaper]{JHEP}
\fi 
\usepackage{amsmath,amssymb}
\usepackage[final]{epsfig}

\newif\ifspinpm 
\spinpmtrue 

\allowdisplaybreaks[3]
\ifom
\input dnmacs
\fi 

\def\omt{{\ifom{{\dn\dnhalf :}}\else%
        {{3\!{\footnotesize$\mathbf{{\frown}\llap{\text{\tiny$\prime$}}}$}%
        {\hbox to -.7ex{\null}\llap{\raise1.3ex\hbox{\tiny{%
        \setbox255=\hbox{$\mathbf{{\smile}}$}%
        \copy255\kern-.7\wd255{\raise.5ex\hbox{$\mathbf{\cdot}$}}}}}}}}\fi}}

\newcommand\skipthis[1]{{}}

\let\oldAE\AE
\renewcommand\AE{{\ifmmode{\text{\it\oldAE}}\else{\oldAE}\fi}}

\newcommand\ct[1]{{\ifuseprd{\em{#1}},\else{\sf {#1}},\fi}}

\chardef\til=`~

\newcommand\Dp{| Dp \rangle}
\newcommand\BZ{\mathbb{Z}}

\providecommand\FIGURE[2][]{\begin{figure}[#1]\begin{center}{#2}\end{center}
                       \end{figure}}
\providecommand\putabstract[1]{\ifuseprd\begin{abstract} {#1} \end{abstract}%
                           \else \abstract{{#1}} \fi}
\providecommand\plb[3]{{Phys.\ Lett.\ B {\bf {#1}}, {#3} ({#2})}}
\providecommand\cqg[3]{{Class.\ Quant.\ Grav.\ {\bf {#1}}, {#3} ({#2})}}
\providecommand\npb[3]{{Nucl.\ Phys.\ {\bf B{#1}}, {#3} ({#2})}}
\providecommand\jhep[3]{{J.\ High Energy Phys.\ {\bf #1}, {#3} ({#2})}}
\providecommand\npps[3]{{Nucl.\ Phys.\ {bf {#1}} Proc.\ Suppl.\ {#3} ({#2})}}

\providecommand\hepth[1]{{\tt hep-th/{#1}}}
\providecommand\hepph[1]{{\tt hep-ph/{#1}}}

\ifuseprd
\begin{document} 
\fi 

\title{Forces between stable non-BPS branes}
\ifuseprd
\author{Steven Corley}\email{scorley@het.brown.edu}
\author{David A. Lowe}\email{lowe@het.brown.edu}
\affiliation{Department of Physics \\ 
       Brown University \\
       Providence, RI \ 02912 \\
       USA}
\else 
\author{Steven Corley\thanks{\tt scorley@het.brown.edu} \ and David A.
Lowe\thanks{\tt lowe@het.brown.edu} \\
Department of Physics\\
Brown University \\
Providence, RI \ 02912\\
USA}
\fi 

\putabstract{
As a step toward constructing realistic brane world models in string
theory, we consider the interactions of a pair of non-BPS branes. We
construct a dyonic generalization of the non-BPS branes first
constructed by Bergman, Gaberdiel and Sen as orbifolds of D-branes
on $T^4/\BZ_2$. The force between a
dyonic brane and an electric brane is computed and
is found to vanish at a nontrivial critical separation. This
equilibrium point is unstable. For smaller separations the branes
coalesce to form a composite dyonic state, while for larger
separations the branes run off to infinity. We suggest generalizations
that will lead to potentials with stable local minima.
}

\preprint{BROWN-HET-1287\ifuseprd,~\else\\\fi}

\ifuseprd
\maketitle
\else
\begin{document}
\fi 

\section{Introduction}

The existence of compact extra dimensions with sizes of order a millimeter
appears to be consistent with all known experiments \cite{AHpheno}. If
true, the fundamental scale for physics may lie in the range $10-100$
TeV, and the hierarchy problem becomes explaining why the size of
these extra dimensions is so large. For this new view to be
consistent, one must postulate that the Standard Model fields are
confined to a hypersurface in this higher dimension geometry -- a
``brane-world'' \cite{extradlist}. This
idea is well-motivated in the context of string theory, where D-branes 
play exactly this role. A wide variety of gauge groups and matter
content can be found as exact string theory compactifications (see
\cite{stringcom,nonbpslist} for some reviews).

However we are still a long way off from reproducing the known
Standard Model (with no other light fields) as an exact
compactification of string theory. A step in this direction was taken
in \cite{Sen0} where stable non-supersymmetric D-brane states were
constructed in orbifolds of Type II string theory.
Similar constructions developing more realistic gauge and matter
contents followed (see \cite{nonbpslist} for a review). However these compactifications
are always accompanied by unwanted light fields associated with
rescaling the sizes of internal dimensions (or the dilaton, which in
turn is related to the size of an 11th direction in the M-theory
viewpoint). These light fields are often referred to as radions.

At the phenomenological level, suggestions for stabilizing 
radion fields have been made in \cite{ahstab,branecrystal}. In
particular, in \cite{branecrystal} we showed the hierarchy problem
could be solved by having a crystal structure in the internal
dimensions, involving a large number of branes. For this to work, the
forces between branes must balance at some finite critical radius, of
order the fundamental length scale.
In this paper we will take a step toward realizing this mechanism as
an exact solution of string theory. 

We will begin by generalizing the non-BPS branes of \cite{Ber1,Senrev}
(see also \cite{L1}) to
carry additional charge with respect to a $p-1$ form field
strength. The interaction potential between such a dyonic brane and a purely
electrically charged brane takes a highly nontrivial form that does
indeed display an extremum at finite brane separation. Unfortunately
for the example constructed here, this extremum is a local
maximum. The branes may either run off to infinite separation, or they 
may coalesce. We conjecture the branes will form a stable composite
dyonic state. Similar bound states have been discussed for pairs of
pure 
electric
case in \cite{lambertsachs}. A supergravity solution for a stack of a
large number of electric or magnetic branes has been constructed in
\cite{gian}. We suggest that the inclusion of other types of brane
charge will lead to a true stable minimum.

To describe the non-BPS branes considered in this paper we employ
the boundary state formalism.  Such methods provide a 
convenient calculational tool for computing the potential between
a pair of D-branes.
Specifically for a pair of D-branes described by the boundary states
$| D_1 \rangle$ and $|D_2 \rangle$ respectively, the potential
between them is given by
\begin{equation}
V_{1-2} = - \langle D_1 | {\cal D} | D_2 \rangle
\label{defineV}
\end{equation}
where ${\cal D}$ is the closed string propagator
\begin{equation}
{\cal D} = \frac{\alpha'}{4 \pi} \int_{|z| \leq 1} \frac{d^2 z}
{|z|^2} z^{L_{0}} \bar{z}^{\tilde{L}_0}.
\end{equation}
In the open string description of D-branes, one would instead have
to compute the 1-loop partition function - or annulus diagram -
with the open string endpoints on either brane.  The advantage
of the boundary state formalism is its universality.
Once a boundary state
is known one need only plug into (\ref{defineV}) to find the
potential between a pair of branes.  In the open string description
one must recompute the mode expansion for the open
string each time one of the D-branes ($|D_{1,(2)} \rangle$) is
changed as well as find the corresponding projection operator
to be inserted in the partition function.
The boundary state formalism has been applied in a variety of
cases to study the properties of non-BPS branes (some reviews
may be found in \cite{boundarystate, diVLic, diVLic2, L5}).

\section{Constructing the boundary state}

The next few subsections will be devoted to the construction
of the boundary states used in this paper for computing the
potential between a pair of non-BPS Dp-branes in type IIB
for p even or type IIA for p odd.  Before discussing
the details let us first pause for a moment to specify the
setup.  Recall that non-BPS branes are in general unstable - they
support tachyonic excitations.  In some cases they can be
stabilized by an appropriate orbifolding.  The relevant
example for us will be to
take the $x^6, x^7, x^8, x^9$ directions to lie on a torus
$T^4$ with the $(p+1)$ directions tangent to the brane
lying in the noncompact directions.  Modding out by 
${\cal I}_4(-1)^{F_L}$ where ${\cal I}_4$ reverses the signs of the $T^4$
coordinates and $F_L$ denotes the contribution to the spacetime
fermion number coming from the left-moving sector of the 
worldsheet removes the tachyon field from the non-winding
modes of the string.
For torus radii $R_6, R_7, R_8, R_9$ all larger than
the critical value $\sqrt{\alpha'/2}$ this is enough to remove
all tachyonic modes from winding string. 
 
In the next subsection we compute the 1-loop partition
function for each of the individual branes that we shall
consider.
This is a necessary step in order to fix various
coefficients in the boundary states, which we then construct
in subsection \ref{ss:boundarystates}.  In section \ref{ss:potential}
we use the definition (\ref{defineV}) to construct
the potential between the two boundary states of interest.

\subsection{1-loop partition function}
\label{ss:1loop}

In this subsection we compute the 1-loop partition function
for the non-BPS branes of interest in this paper, specifically
one carrying charge associated to the twisted sector (p+1)
form RR potential and the other carrying charges associated
with the twisted sector (p+1) and (p-1) form RR potentials.
The former is a special case of the latter so we begin with
it.

The inclusion of lower brane charge can be accomplished by
turning on a constant $B_{\mu \nu}$ (or equivalently a
constant $F_{\mu \nu}$) field.  The resulting sigma
model action is given by
\begin{equation}
S_{open} = -\frac{1}{4 \pi \alpha'} \int d^{2} \sigma ( \eta_{M N}
\eta^{AB} \partial_{A} X^{M} \partial_{B} X^{N} + \epsilon^{AB}
B_{MN} \partial_{A} X^{M} \partial_{B} X^{N} )
\label{openaction}
\end{equation}
where $\epsilon^{AB}$ is antisymmetric, $\epsilon^{01} = 1$, and
we follow the metric conventions
$\eta_{MN} = diag(-1,1,...,1)$ and $\eta_{AB} = diag(-1,1)$.
Our index notation will be to use $M,N,...$ indices for 10 dimensional
spacetime indices which we decompose as $M = (\mu, i, a)$ where
$\mu$ runs over the brane coordinates $\mu = 0,...,p$, $i$ runs over
the remaining noncompact dimensions $i = p+1,...,5$, and $a$ runs
over the $T^4$ coordinates $a=6,...,9$.  Also we have used
$A, B,...$ to denote worldsheet indices.

For the constant $B$ field to give rise to codimension 2 lower
brane charge we must have rank 2 $B$ field.  We therefore take
$B_{12} = f = -B_{21}$ with all other components of $B$ set to
zero.  The boundary conditions on the open string endpoints 
following from the above action are then
\begin{eqnarray}
\partial_{\sigma} X^{\mu}|_{\sigma = 0,\pi} & = & 0, 
\,\,\,\, \mu \neq 1,2 \nonumber \\
(\partial_{\sigma} X^1 - f \partial_{\tau} X^2)|_{\sigma = 0,\pi} & = & 0
\nonumber \\
(\partial_{\sigma} X^2 + f \partial_{\tau} X^1)|_{\sigma = 0,\pi} & = & 0
\nonumber \\
X^{M} |_{\sigma = 0,\pi} & = & x^{M}, \,\,\,\, M=i,a.
\label{endpoints}
\end{eqnarray}

The worldsheet fermions are handled in the usual way with one
exception.  The boundary conditions for the $M \neq 1,2$ fermions
are the standard ones, namely the right-moving, $\psi_+$,
and left-moving, $\psi_-$, fermions are related by
\begin{equation}
\psi^{M}_+ |_{\sigma = 0} = \psi^{M}_- |_{\sigma = 0}, \, \, \,
\psi^{M}_+ |_{\sigma = \pi} = \psi^{M}_- |_{\sigma = \pi}
\end{equation}
in the Ramond sector and by
\begin{equation}
\psi^{M}_+|_{\sigma = 0} = \psi^{M}_- |_{\sigma = 0}, \, \, \,
\psi^{M}_+|_{\sigma = \pi} = - \psi^{M}_- |_{\sigma = \pi}
\end{equation}
in the Neveu-Schwarz sector.  In order to
preserve worldsheet supersymmetry however the $M=1,2$ fermion
boundary conditions must be modified to
\begin{eqnarray}
(\psi^{1}_+ - \psi^{1}_-)|_{\sigma = 0} & = & f(\psi^{2}_+ + \psi^{2}_-)
|_{\sigma = 0}, \nonumber \\
(\psi^{2}_+ \mp \psi^{2}_-)|_{\sigma = \pi} & = & f(\psi^{1}_+ \pm \psi^{1}_-)
|_{\sigma = \pi},
\end{eqnarray}
where the upper(lower) sign in the second equation applies in the
R(NS) sector.

The mode expansions for the worldsheet fields subject to the
above boundary conditions are obtained
in the standard way, for the details see \cite{ACNY1} for the
boundary conditions involving an $f$ and eg. \cite{Polchinski12}
for the other boundary conditions.  We find the following
expansions:
\begin{eqnarray}
Z & = & z + 2 \alpha' \frac{p (\tau - i f \sigma)}{1 + f^2} + i \sqrt{2
  \alpha'}
\sum^{\infty}_{n=1} \frac{1}{n} \Bigl( a_{n} e^{-i n \tau} \cos(n \sigma +
  {\cal \phi}) \nonumber \\
& - & b^{\dagger}_{n} e^{i n \tau} \cos(n \sigma - {\cal \phi}) \Bigr)
\label{Zexp} \\
X^{\mu} & = & x^{\mu} + 2 \alpha' p^{\mu} \tau + i \sqrt{2 \alpha'}
\sum^{\infty}_{n=-\infty,\neq 0} \frac{1}{n} a^{\mu}_{n} e^{-i n \tau}
\cos{n \sigma}, \,\,\,\, \mu \neq 1,2 \\
X^{M} & = & x^{M}_{1} + \frac{\sigma}{\pi} (x^{M}_2 - x^{M}_1)
+ i \sqrt{2 \alpha'}
\sum^{\infty}_{n=-\infty,\neq 0} \frac{1}{n} a^{M}_{n} e^{-i n \tau}
\cos{n \sigma}, \,\,\,\, M = i,a
\end{eqnarray}
for the worldsheet bosons and
\begin{eqnarray}
\psi_+ & = & \sqrt{2 \alpha'} \frac{1-if}{\sqrt{1 + f^2}} \sum_{n} c_{n} 
e^{-in(\tau + \sigma)}, \,\,\,
\psi_- = \sqrt{2 \alpha'} \frac{1+if}{\sqrt{1 + f^2}} \sum_{n} c_{n} 
e^{-in(\tau - \sigma)}, \\
\psi^{M}_{+} & = & \sqrt{\alpha'} \sum_{n} c^{M}_n e^{-in(\tau + \sigma)},
\,\,\, \psi^{M}_{-} = \sqrt{\alpha'} \sum_{n}  c^{M}_n
e^{-in(\tau - \sigma)},
\,\,\, M \neq 1,2
\end{eqnarray} 
for the worldsheet fermions.
The $1,2$ fields are given in terms of the above fields through
the definitions
\begin{eqnarray}
Z & = & \frac{1}{\sqrt{2}} (X^{1} + i X^{2}), \\
\psi_{\pm} & = & \psi^{1}_{\pm} + i \psi^{2}_{\pm}.
\end{eqnarray}
The phase ${\cal \phi}$ in the $Z$ mode expansion (\ref{Zexp}) is given
in terms of $f$ by
\begin{equation}
{\cal \phi} = \frac{\pi}{2} - \tan^{-1}(1/f).
\end{equation}
The index sum in the fermion expansions is over half-integers in 
the NS sector and integers in the R sector.  The (anti-)commutation
relations of the worldsheet fields imply the following mode
(anti-)commutation relations
\begin{eqnarray}
[z,\bar{p}] & = & i, \,\, 
[a_n, a^{\dagger}_m]  =  n \delta_{n-m}, \,\, [b_n, b^{\dagger}_m] =
n \delta_{n-m}, \\
\, [x^{\mu}, p^{\nu} ] & = & i \eta^{\mu \nu}, \,\, \mu, \nu \neq 1,2, \,\,\,
[a^{M}_n, a^{N}_m]  =  n \eta^{MN} \delta_{m+n}, \,\, M,N \neq 1,2,
\label{comm} \\
\{c_n, c^{\dagger}_m\} & = & \delta_{n-m}, \,\,\, \{c^{M}_n, c^{N}_m \}
= \eta^{MN} \delta_{m+n}, \,\, M,N \neq 1,2
\end{eqnarray}
with all other (anti-)commutation relations vanishing.

{}From the above mode expansions it is straightforward to 
construct the Virasoro
generators and in particular one finds for $L_{0}$,
\begin{equation}
L_{0} = \alpha' p^{\mu} g_{\mu \nu} p^{\nu} + \frac{1}{4 \pi^2
 \alpha'}
(x_2 - x_1)^2 + \sum_{n=1}^{\infty} ( \eta_{MN} a^{M}_{-n} a^{N}_{n} 
+ n \eta_{MN} c^{M}_{-n} c^{N}_{n} )
\label{L0}
\end{equation}
where we have made the definitions
\begin{eqnarray}
 a^{1}_{n} & = & (a_n - a^{\dagger}_n)/\sqrt{2}, \,\,
a^{1}_{-n} = (a_n + a^{\dagger}_n)/\sqrt{2}, \\
a^{2}_n & = & (b_n - b^{\dagger}_n)/\sqrt{2}, \,\,
a^{2}_{-n} = (b_n + b^{\dagger}_n)/\sqrt{2}, \\
p^1 & = & (p + \bar{p})/\sqrt{2}, \,\, p^{2} = (p - \bar{p})/\sqrt{2},
\\
c_n & = & (c^{1}_n + i c^{2}_n)/\sqrt{2}, \,\,
c^{\dagger}_n = (c^{1}_{-n} + i c^{2}_{-n})/\sqrt{2}, \\
g_{\mu \nu} & = & \mbox{diag}(-1, 1/(1+f^2),1/(1+f^2),1, \cdots, 1).
\end{eqnarray}
These $a^{1,2}$ oscillators now satisfy the commutation relations
in (\ref{comm}) for $M,N=1,2$.

The partition function is given by
\begin{equation}
Z = \int^{\infty}_{0} dt \, \frac{1}{2t} \mbox{Tr}_{NS-R}
\Bigl( {\cal P} e^{-2 \pi t L_0} \Bigr)
\label{partition}
\end{equation}
where ${\cal P}$ is a projection operator.  Recall that
a single non-BPS brane has two Chan-Paton factors, the
identity ${\cal I}$ and Pauli matrix $\sigma_1$ (the
other possible Chan-Paton factors $\sigma_2$ and $\sigma_3$
are projected out in the construction of the non-BPS brane
from a brane-anti-brane pair, see eg. \cite{Senrev}).  Each Chan-Paton factor
has its own projection operator.  For the
orbifold that we are considering, $T^4 /{\cal I}_4 (-)^{F_L}$,
these projections have been worked out \cite{Ber1,Senrev} and are
given by
\begin{equation}
{\cal P}_{I,\sigma_1} = \frac{1 \pm (-)^{F}}{2} \frac{1 \pm {\cal I}_4
  (-)^{F_L}}{2}
\end{equation}
where the upper (lower) sign corresponds to the $I$ ($\sigma_1$)
Chan-Paton factor.  The partition function is then a sum
of partition functions for the $I$ and $\sigma_1$ open
string sectors.  The trace appearing in each of these
sectors however is over the same set of states so that
the projection operator appearing in (\ref{partition}) is
just ${\cal P} = {\cal P}_I + {\cal P}_{\sigma_1}$, which is
simply
\begin{equation}
{\cal P} = \frac{1 + (-)^F {\cal I}_4 (-)^{F_L}}{2}.
\end{equation}
Evaluating the partition function is now a straightforward task
given all the data accumulated previously.  The end result is
\begin{eqnarray}
Z & = & \int^{\infty}_0 dt \, \frac{1}{2t} \frac{V_{p+1}}{(2 \pi)^{p+1}}
(1 + f^2) (2 \alpha' t)^{-(1+p)/2} e^{- 
\frac{({\bf x} - {\bf y})^2}{2 \pi \alpha'} t}
\biggl(\prod_{j=6}^9 \Bigl( \sum_{n_j = -\infty}^{\infty}
e^{- \frac{2 \pi}{\alpha'} (n_j R_j)^2 t} \Bigr) \nonumber \\
& \cdot & 
\frac{(f_3(e^{- \pi t}))^8 - (f_2(e^{- \pi t}))^8}{(f_1(e^{-\pi t}))^8}
-4 \Bigl(\frac{f_3(e^{- \pi t}) f_4(e^{- \pi t})}{f_1(e^{- \pi t}) 
f_2(e^{- \pi t})} \Bigr)^4 \biggr)
\label{parteval}
\end{eqnarray}
where the functions $f_i$ are defined as
\begin{eqnarray}
f_1(q) & = & q^{1/12} \prod_{n=1}^{\infty} (1 - q^{2n}) \\
f_2(q) & = & \sqrt{2} q^{1/12} \prod_{n=1}^{\infty} (1 + q^{2n}) \\
f_3(q) & = & q^{-1/24} \prod_{n=1}^{\infty} (1 + q^{2n-1}) \\
f_4(q) & = & q^{-1/24} \prod_{n=1}^{\infty} (1 - q^{2n-1}). 
\end{eqnarray}
In obtaining the result (\ref{parteval}) we have used the
covariant formalism.  In particular the result (\ref{parteval})
includes the ghost contribution.  Since this contribution
is independent of the background $B$-field we have not
bothered to give the details, which can be found in 
eg. \cite{CLNY2,PolCai}.   

The dependence of the partition function (\ref{parteval})
on the background $B_{MN}$ field is quite simple in that
$f$ only enters in an overall multiplicative factor.
Taking $f \rightarrow 0$ yields the partition function
for a non-BPS in vanishing $B_{MN}$ field.  This
partition function agrees with that computed elsewhere
\cite{GabSen} and serves as a useful check on our 
calculations.

\subsection{Construction of the Boundary State}
\label{ss:boundarystates}

The boundary state description of D-branes has been widely
used so we shall limit our discussion here to primarily
listing the relevant formulae.  In particular
the construction of boundary states in the presence of external
fields has been discussed in \cite{L3, Kam1, Kam2}.   Some useful reviews on
the subject are \cite{boundarystate,diVLic,diVLic2,L5}.  

The two
main problems are to determine the boundary conditions
satisfied by the state and to fix the appropriate GSO
projection for the orbifold under consideration.  The
first problem is easily handled by converting the open
string boundary conditions in the previous section
to the closed string boundary conditions via the procedure
reviewed in \cite{diVLic}.  The result is
\begin{eqnarray}
\partial_{\tau} X^{\mu} (0,\sigma) \Dp & = & 0, \,\,\,\,
\mu = 0,3,...,p \\
(\partial_{\tau} X^{1}(0,\sigma) - f \partial_{\sigma}
X^{2}(0,\sigma)) \Dp & = & 0 \\
(\partial_{\tau} X^{2}(0,\sigma) + f \partial_{\sigma}
X^{1}(0,\sigma)) \Dp & = & 0 \\
X^{M}(0,\sigma) \Dp & = & 0, \,\,\, M=(p+1),...,9
\end{eqnarray}
for the bosonic fields and
\begin{equation}
\psi^{M}_{-}(0,\sigma) = i \eta S^{M}_{\hspace{0.5em} N}
\psi^{N}_{+}(0,\sigma)
\end{equation}
for the fermionic fields where the matrix $S^{M}_{\hspace{0.5em} N}$
is block diagonal and is the identity in the $M,N=0,3,...,9$ block
and 
\begin{equation}
S^{M}_{\hspace{0.5em} N} = \frac{1}{1+f^2} \left(
\begin{array}{cc}
1-f^2 & -2f \\
2f & 1-f^2 
\end{array} \right)
\end{equation}
in the 1,2 block.  The constant $\eta$ can be $\pm 1$ and both
possibilities arise in the final boundary state. 

Solving these equations is straightforward given the closed
string mode expansions.  The latter for the bosonic string coordinates
is given by
\begin{eqnarray}
X^{M}(\tau,\sigma) & = & \hat{x}^M 2 + \alpha' \Bigl( \hat{p}^{M}_L
(\tau + \sigma) + \hat{p}^{M}_R (\tau - \sigma) \Bigr) \nonumber \\
& + & i \sqrt{\alpha'/2} \sum_{n \in {\BZ}, \neq 0} \frac{1}{n}
\Bigl(\alpha^{M}_n e^{-i 2 n(\tau - \sigma)} +
\tilde{\alpha}^{M}_n e^{-i 2 n(\tau + \sigma)} \Bigr)
\end{eqnarray}
in the untwisted sector where
\begin{equation}
\hat{p}^{M}_L = \frac{1}{2} \Bigl(\frac{n_M}{R^M}
+ \frac{m_M R^M}{\alpha'} \Bigr), \,\,\,
\hat{p}^{M}_R = \frac{1}{2} \Bigl(\frac{n_M}{R^M}
- \frac{m_M R^M}{\alpha'} \Bigr)
\end{equation}
in the compact directions and $\hat{p}^{M}_L = \hat{p}^{M}_R
= \hat{p}^{M}$ in the noncompact directions.
In the twisted sector the mode expansion in the
compact directions is given by
\begin{equation}
X^{a}(\tau,\sigma) = x^{a} +
i \sqrt{\alpha'/2} \sum_{n \in {\BZ} + 1/2} \frac{1}{n}
\Bigl(\alpha^{a}_n e^{-i 2 n(\tau - \sigma)} +
\tilde{\alpha}^{a}_n e^{-i 2 n(\tau + \sigma)} \Bigr)
\end{equation}
assuming that the branes are located at the one of the orbifold fixed
planes $x^{a} = 0, \, \pi R^{a}$.
The fermionic mode expansions are given by
\begin{eqnarray}
\psi^{M}_- & = & \sqrt{2 \alpha'} \sum_{t} \psi^{M}_t e^{-i 2 t (\tau
- \sigma)} \\
\psi^{M}_+ & = & \sqrt{2 \alpha'} \sum_{t} \tilde{\psi}^{M}_t 
e^{-i 2 t (\tau + \sigma)}
\end{eqnarray}
where the index $t$ satisfies
\begin{equation}
t \in \left\{
\begin{array}{r@{\quad:\quad}l}
{\BZ} + 1/2 & \mbox{untwisted NS or twisted R} \\
{\BZ} & \mbox{untwisted R or twisted NS}.
\end{array} \right.
\end{equation}
and the twisted boundary conditions only apply in the
compactified directions, $M=a$.

Solving for the boundary states yields
\begin{equation}
|Dp,f,x^{i},x^{a},\eta \rangle_{U} = {\cal N}_{1,f} \int \Bigl(\prod_{i} dk_i
e^{i k_{j} x^{j}} \Bigr) \Bigl(\prod_{a} \sum_{m_a} 
e^{i m_a x^a /R^a} \Bigr) |Dp,f,k,m \rangle_{X,U} 
|Dp,f,\eta \rangle_{\psi,U}
\end{equation}
in the untwisted sector where the $X$ and $\psi$
pieces of the state are given by
\begin{eqnarray}
|Dp,f,k,m \rangle_{X,U} & = & \exp \Bigl( - \sum_{n > 0}
\frac{1}{n} \alpha^{M}_{-n} S_{MN} \tilde{\alpha}^{N}_{-n}
\Bigr) |0_{\alpha},k_{i},m_{a} \rangle \label{Xpart} \\
|Dp,f,\eta \rangle_{\psi,U} & = & \exp \Bigl( i \eta
\sum_{t>0} \psi^{M}_{-t} S_{MN} \tilde{\psi}^{N}_{-t} \Bigr)
|Dp,f,\eta \rangle^{(0)}_{\psi}
\label{psipart}
\end{eqnarray}
respectively where the indices are as appropriate for
the untwisted $R$ and $NS$ sectors.  We will discuss
the zero mode contribution $|Dp,\eta \rangle^{(0)}_{\psi}$ shortly.
Similarly for the twisted sector we find
\begin{equation}
|Dp,f,x^{i},x^{a}, \eta \rangle_{T} = {\cal N}_{2,f} \int \Bigl(\prod_{i} dk_i
e^{i k_{j} x^{j}} \Bigr) |Dp,f,k \rangle_{X,T} 
|Dp,f,\eta \rangle_{\psi,T}
\end{equation}
where the twisted sector matter states $ |Dp,f,k \rangle_{X,T}$ 
and $|Dp,f,\eta \rangle_{\psi,T}$ are exactly as in (\ref{Xpart})
and (\ref{psipart}) with the appropriate changes in the
index summations.

The zero mode contribution to the $\psi$ boundary state is
not difficult to find, but as we shall see later the only
non-trivial contribution to it that we need comes from the
twisted $R$ sector.  In this sector only the $M=(\mu,i)$
worldsheet fermions have zero modes.  To simplify the notation
we let $\alpha, \beta,.. = 0,...,5$.  A convenient representation
of the zero mode anticommutation relations is given by \cite{eyras}
\begin{eqnarray}
\psi^{\alpha}_0 |a,\tilde{b} \rangle & = & \frac{1}{\sqrt{2}}
(\gamma^{\alpha})^{a}_{\hspace{0.4em} c} \, \delta^{b}_{\hspace{0.4em} d}
|c,\tilde{d} \rangle \\
\tilde{\psi}^{\beta}_0 |a,\tilde{b} \rangle & = & \frac{1}{\sqrt{2}}
\gamma^{a}_{\hspace{0.4em} c} \, (\gamma^{\beta})^{b}_{\hspace{0.4em} d}
|c,\tilde{d} \rangle
\end{eqnarray}
where the $\gamma$ matrices satisfy the $SO(1,5)$ Clifford algebra
$\{ \gamma^{\alpha}, \gamma^{\beta} \} = 2 \eta^{\alpha \beta}$
and $\gamma = - \gamma^{0} \gamma^{1}\cdots \gamma^{5}$.  A simple
calculation then yields
\begin{equation}
|Dp,f,\eta \rangle^{(0)}_{\psi,T,R} = 
\Bigl[C \gamma^0 \gamma^3 \cdots \gamma^p
\frac{1 + (1/f) \gamma^1 \gamma^2}{\sqrt{1+1/f^2}} 
\frac{1 + i \eta \gamma}{1+i \eta} \Bigr]_{ab} | a, \tilde{b} \rangle
\label{zeromode}
\end{equation}
where we have taken an arbitrary normalization (the overall
normalization will be fixed below). 

We have so far ignored the ghost contributions to the boundary
states listed above.  Since we are using the covariant formalism
however it is crucial that we include these terms.  As it turns
out though the ghost boundary state is independent of the orbifold
that we are taking, i.e., it is the same state as derived for
the flat Minkowski background in \cite{CLNY2,PolCai}.  The relevant
formulae are nicely collected in the review \cite{diVLic} and
we shall not bother to rewrite everything here.    

To construct the boundary state corresponding to the non-BPS
brane that we want we must find the correct GSO projection
corresponding to the orbifold configuration that we have
taken.  This has already been done \cite{Ber1}.  In the
untwisted sector one has the usual type IIA/B GSO projection.
For a non-BPS brane this leaves the $NS-NS$ sector part of the
untwisted state but removes the $R-R$ piece as it
has the ``wrong'' worldvolume dimension.  On the twisted
sector side the $NS-NS$ part of the state is projected out
while the $R-R$ sector piece remains.  The resulting boundary state is
\begin{eqnarray}
|Bp,f,x^i,x^a,\epsilon \rangle & = & \frac{1}{2} \Bigl(|Dp,f,x^i,x^a,+
\rangle_{NSNS,U} - |Dp,f,x^i,x^a,- \rangle_{NSNS,U} \Bigr) 
\nonumber \\
& + & \frac{\epsilon}{2} \Bigl(
|Dp,f,x^i,x^a,+ \rangle_{RR,T} + |Dp,f,x^i,x^a,- \rangle_{RR,T}
\Bigr)
\label{Bp}
\end{eqnarray} 
where $\epsilon$ is $\pm 1$ corresponding to a (anti-)brane\footnote{
Note that the $\epsilon$ appearing in the definition of the 
boundary state $|Bp \rangle$ is not to be confused with the
$\eta$ used in constructing the $|Dp \rangle$ states.  The 
boundary state $|Bp \rangle$ in fact contains both
$\eta = 1$ and $\eta=-1$  $|Dp \rangle$ states in its definition
in (\ref{Bp}).}. 

The final step in the construction of the boundary state
is to compute the normalization factors ${\cal N}_{1,f}$ and
${\cal N}_{2,f}$.  This is done by computing the one-loop
partition function for open strings on the non-BPS
brane using the above boundary state and comparing to
the open string computation of the previous section.
Given the closed string propagator
\begin{equation}
{\cal D}  = \frac{\alpha'}{4 \pi} \int_{|z| \leq 1} 
\frac{d^2 z}{|z|^2} z^{L_0 - a} \bar{z}^{\tilde{L}_0 - a},
\label{propagator}
\end{equation}
where the normal ordering constant $a$ is $1/2$ in the
untwisted $NS-NS$ sector and 0 otherwise,
then the one loop partition function is given in
terms of the boundary state by
\begin{equation}
Z = \langle Bp,f,x^i,x^a,\epsilon | {\cal D}
| Bp, f, x^i, x^a, \epsilon \rangle.
\label{bstateZ}
\end{equation}
The matter contribution to the Virasoro generator $L_0$
is given by
\begin{equation}
L_0 = \alpha' \hat{p}_{L}^2 + \sum_{n>0} \alpha_{-n} \cdot \alpha_n
+ \frac{1}{2} \sum_{r >0} r \psi_{-r} \psi_r
\end{equation}
with a similar expression for $\tilde{L}_0$.  The indices here
differ in different sectors ($NS$ versus $R$ and twisted versus
untwisted) as discussed previously.

Evaluation of the partition function (\ref{bstateZ}) is now a
straightforward task modulo one subtlety involving the zero
modes.  The point is simply that naive evaluation of the inner
product 
$_{\psi,T,R}^{(0)} \langle Dp,f,\eta_1|Dp,f,\eta_2
\rangle^{(0)}_{\psi,T,R}$ (in which we really mean not just the 
state (\ref{zeromode}) but also the ghost zero mode contribution
as well given in eg. \cite{diVLic}) would yield a divergent
result.  One can however define this inner product \cite{Billo} through
the regularization
\begin{equation}
^{(0)}_{\psi,T,R} \langle Dp,f,\eta_1|Dp,f,\eta_2 
\rangle^{(0)}_{\psi,T,R} = \lim_{x \to 1} \;
^{(0)}_{\psi,T,R} \langle Dp,f,\eta_1|x^{2(F_0 + G_0)}|Dp,f,\eta_2 
\rangle^{(0)}_{\psi,T,R}
\end{equation}
where $F_0$ and $G_0$ are the zero mode contributions to the
fermion and superghost number operators.  The details of
the regularization can be found in \cite{Billo}.  With
this regularization we find
\begin{equation}
^{(0)}_{\psi,T,R} \langle Dp,f,\eta_1|Dp,f,\eta_2 
\rangle^{(0)}_{\psi,T,R} = -4 \delta_{\eta_1 \eta_2 -1}
\end{equation}
where the right-hand-side is independent of $f$.  In the
next section we shall require this inner product in the
case in which one of the boundary states has vanishing
$f$, the result is
\begin{equation}
^{(0)}_{\psi,T,R} \langle Dp,0,\eta_1|Dp,f,\eta_2 
\rangle^{(0)}_{\psi,T,R} = -\frac{4}{\sqrt{1+f^2}} \delta_{\eta_1 \eta_2 -1}
\end{equation}
which is not independent of $f$ and will play an important role later.

Finally one can determine the normalization constants
${\cal N}_{1,f}$ and ${\cal N}_{2,f}$ by comparing the
boundary state computation of the partition function
(\ref{bstateZ}) to the open string evaluation (\ref{parteval}).
We find 
\begin{eqnarray}
({\cal N}_{1,f})^{2} & = & \frac{1+f^2}{2^{p+5} \pi^{p+2}
(\alpha')^{p-3} R_6 \cdots R_9} \\
({\cal N}_{2,f})^{2} & = & \frac{1+f^2}{2^{p+1} \pi^{p+1}
(\alpha')^{p-1}}.
\end{eqnarray}
Similarly for ${\cal N}_{1,0}$ and ${\cal N}_{2,0}$
one simply takes $f=0$ in the above expressions.
In obtaining these results we have used the identity
\begin{equation}
\sum_{m \in {\BZ}} e^{-(\pi/2) \alpha' t (m/R)^2}
= \sqrt{\frac{2}{\alpha' t}} R \sum_{n \in {\BZ}}
e^{-(2 \pi/\alpha' t)(nR)^2}
\end{equation}
as well as the modular transformation properties of the $f_i$'s
\begin{eqnarray}
f_1 (e^{-\pi/t}) & = & \sqrt{t} f_1 (e^{-\pi t}), \,\,\,
f_2(e^{-\pi/t}) = f_4 (e^{- \pi t}) \\
f_3 (e^{- \pi/t}) & = & f_3 (e^{- \pi t}), \,\,\,
f_4 (e^{-\pi/t}) = f_2 (e^{- \pi t}).
\end{eqnarray}

\section{Computation of the potential}
\label{ss:potential}

We now have all the ingredients to compute the potential
between the pair of non-BPS branes of interest, ie., both 
charged under the twisted
sector RR $p+1$ form potential with only one also charged
under the twisted sector RR $p-1$ form potential. 
Similar considerations for the interactions between 
non-BPS D-particles in type I string theory can be found
in \cite{L2}. 
The potential between the two is evaluated using
the boundary states through the expression 
(\ref{defineV}).  Specifically for branes located
at $x^{i}$ and $y^{i}$ in the noncompact transverse
directions (and located at the same orbifold fixed
plane in the compact dimensions) we find
\begin{eqnarray}
V & = & -  \langle Bp,0,x^i,x^a,\epsilon_0 | {\cal D}
| Bp, f, y^i, x^a, \epsilon_1 \rangle \nonumber \\
& = & - 2 \frac{V_{p+1}}{(2 \pi)^{p+1}} (2 \alpha')^{-(p+1)/2}
\sqrt{1+f^2} \int_{0}^{\infty} dt \, t^{(p-5)/2} e^{-({\bf x}
-{\bf y})^2/(2 \pi \alpha' t)} \nonumber \\
& \cdot & \biggl( 
\frac{(\alpha'/2)^2}{R_6 R_7 R_8 R_9} \Bigl( \prod_{a=6}^{9}
\frac{f_1(q_a) f_3(q_a)}{f_2(q_a) f_4(q_a)} \Bigr)
\frac{(f_3(q))^6 |f_3(q,\nu)|^2 - (f_4(q))^6 |f_4(q,\nu)|^2}
{(f_1(q))^6 |f_1(q,\nu)|^2} \nonumber \\
& - & \frac{\epsilon_0 \epsilon_1}
{\sqrt{1+f^2}} \frac{(f_2(q))^2 |f_2(q,\nu)|^2 (f_3(q))^4}
{(f_1(q))^2 |f_1(q,\nu)|^2 (f_4(q))^2} \biggr)
\label{potential}
\end{eqnarray}
where we have defined the $f_i$ functions with two arguments as
\begin{eqnarray}
|f_1(q,\nu)|^2 & = & q^{1/6} \prod_{n=1}^{\infty} (1 - e^{i 2 \pi \nu}
q^{2n})(1 - e^{- i 2 \pi \nu} q^{2n}) \\
|f_2(q,\nu)|^2 & = & 2 q^{1/6} \prod_{n=1}^{\infty} (1 + e^{i 2 \pi \nu}
q^{2n})(1 + e^{- i 2 \pi \nu} q^{2n}) \\
|f_3(q,\nu)|^2 & = & q^{-1/12} \prod_{n=1}^{\infty} (1 + e^{i 2 \pi \nu}
q^{2n-1})(1 + e^{- i 2 \pi \nu} q^{2n-1}) \\
|f_4(q,\nu)|^2 & = & q^{-1/12} \prod_{n=1}^{\infty} (1 - e^{i 2 \pi \nu}
q^{2n-1})(1 - e^{- i 2 \pi \nu} q^{2n-1})
\end{eqnarray}
and the various arguments of the $f_i$'s are defined as
\begin{eqnarray}
q & = & e^{- \pi t}, \,\,\, q_a = e^{-\pi \alpha' t/(2 R_a)^2} \\
\nu & = & \frac{1}{\pi} \tan^{-1} f.
\end{eqnarray}
In getting the result (\ref{potential}) we have also used the
identity
\begin{equation}
\sum_{n \in {\BZ}} q^{n^2} = \sqrt{2} \frac{f_1(q) f_3(q)}
{f_2(q) f_4(q)}.
\end{equation}

Although we have not been able to evaluate the integral
in the potential (\ref{potential}) analytically, there are a few useful
analytic limits that one can extract.  The first and
most trivial point is to check that $V$ reduces to the potential
evaluated in \cite{GabSen} which should just be the $f \rightarrow 0$
limit of (\ref{potential}).  Indeed it is straightforward
to show that the two expressions agree. 

A more interesting result is to note that in the limit of
large separation between the branes, i.e., large $({\bf x}
- {\bf y})^2/\alpha'$, the integral will be dominated
by large $t$.  Hence we can determine whether the potential
will be attractive or repulsive at large distances by evaluating
the sign of the integrand at large $t$.  In particular the
quantity in parenthesis in (\ref{potential}) reduces in the
large $t$ limit to just
\begin{equation}
\frac{1}{\xi} \Bigl(3 + \frac{1-f^2}{1+f^2} \Bigr) - \frac{4}{\sqrt{1 + f^2}}
\label{Vsign}
\end{equation}
where we have defined the dimensionless number $\xi$ as
\begin{equation}
\xi = \frac{R_6 R_7 R_8 R_9}{(\alpha'/2)^2}.
\end{equation}
Hence for large separation between the branes we should
find an attractive potential when (\ref{Vsign}) is positive
and a repulsive potential when it's negative.  After some
algebra it is easy to find the boundary between these
two cases, i.e., vanishing asymptotic potential, and it satisfies
\begin{equation}
f_{crit}^2 = 2 \sqrt{\xi^2 - 1} \Bigl( \sqrt{\xi^2 - 1}
+ \xi \Bigr).
\end{equation}

Recall that the open string tachyon on the non-BPS branes is
only projected out for radii $R_a > \sqrt{\alpha'/2}$, hence
the smallest value of $\xi$ compatible with the absence of the
tachyon is $\xi = 1$.  For vanishing asymptotic potential
this corresponds to $f=0$.  In fact \cite{GabSen} showed
in this case that the potential is identically zero
for all brane separations.  For a given $\xi > 1$ 
we then find an asymptotically repulsive potential for $f< f_{crit}$ and
an attractive one for $f > f_{crit}$.  

The last piece of analytical data that we can extract is to 
reverse the above argument, namely for small brane separations,
$r \ll \sqrt{\alpha'}$, the integral (\ref{potential}) should
be dominated by small $t$.  Hence, as before, we can determine
whether the potential will be attractive or repulsive at short
distances.  The term in parentheses in (\ref{potential}) reduces
in the small $t$ limit to
\begin{equation}
\frac{t}{2} \frac{\sin(\pi \nu)}{\sinh(\pi \nu/t)} \Bigl( e^{2 \pi \nu/t} - 4 
e^{\pi \nu /t} +2 - 4 e^{- \pi \nu/t} + e^{- 2 \pi \nu/t}
+ \sum_{j=6}^{9} e^{- \pi (R^{2}_j /(\alpha'/2) - 1)/t} \Bigr).
\label{smallt}
\end{equation}
The important point to note about this expansion, when combined
with the $r^2 =({\bf x} - {\bf y})^2$ dependent
exponential factor in (\ref{potential}), is that it diverges
for small enough $r$!  Specifically if the condition
\begin{equation}
r^2 < r_{crit}^2=2 \pi^2 \alpha' \max( \nu, 1 - \nu - R^{2}_i/(\alpha'/2))
\label{rbound}
\end{equation}
holds, then the integral will diverge to minus infinity. On the other
hand, we may perform an analytic continuation from large $r$ to
define the potential for $r$ smaller than this value. The contribution 
from the small $t$ part of the integral then is proportional to\footnote{We note the
values of $p$ of relevance for us are $p=3$ in Type IIA and $p=2$ and
$p=4$ in Type IIB. The case $p=4$ is rather problematic in six
non-compact dimensions because the potential increases linearly at
long distance.}
\begin{equation}
\label{critpot}
 (r^2 -r_{crit}^2)^{(p-1)/2}~.
\end{equation}

\FIGURE{\epsfig{file=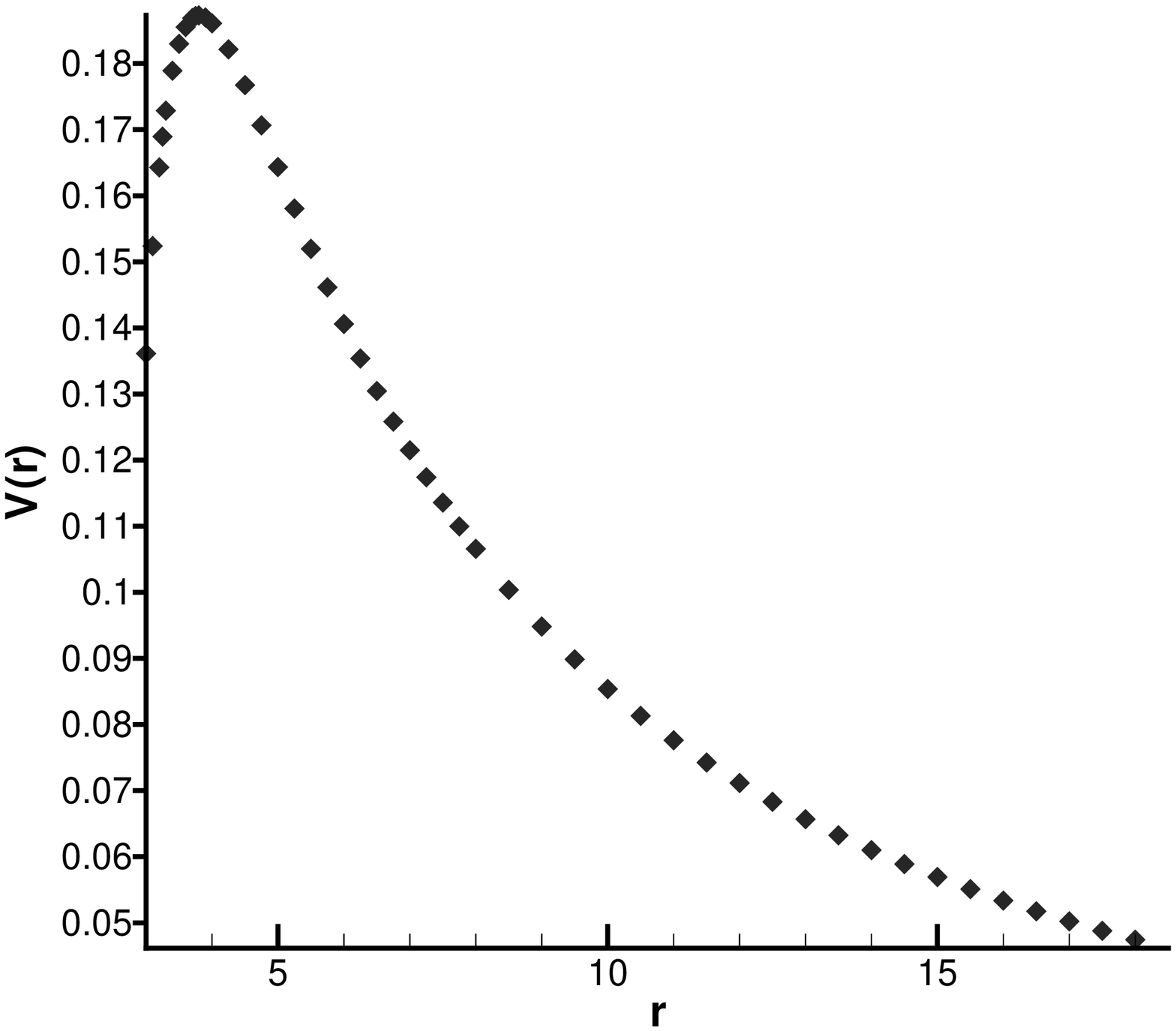,width=10cm}\caption{Brane potential
    for $R_6, \cdots, R_9=\sqrt{2\alpha'}$, $f=10$ and $p=2$.}\label{fone}}

The potential therefore behaves in a similar way to the
brane-anti-brane potential computed in \cite{BS}. 
The potential (\ref{critpot})
is finite at the critical value of $r$ (\ref{rbound}) but
then becomes complex, indicating inelastic modes are opening up. At
the critical separation, the interbrane force diverges for $p<3$,
becoming infinitely attractive. 
At this point, an open string mode running between the different branes
becomes massless, as can be seen from the expansion in (\ref{smallt}).
Specifically one can
view the potential as coming from a one 
loop open string diagram without any external strings,
i.e., just the partition function as in (\ref{partition})
(although with a different GSO projection for the open
strings stretching between the branes).  The coefficients
of the $1/t$ terms in the exponentials in (\ref{smallt})
(including the $r^2$ term) are then just the masses of 
the open string states stretching between the branes.
A mode analysis of these open strings confirms this
expectation.

\FIGURE{\epsfig{file=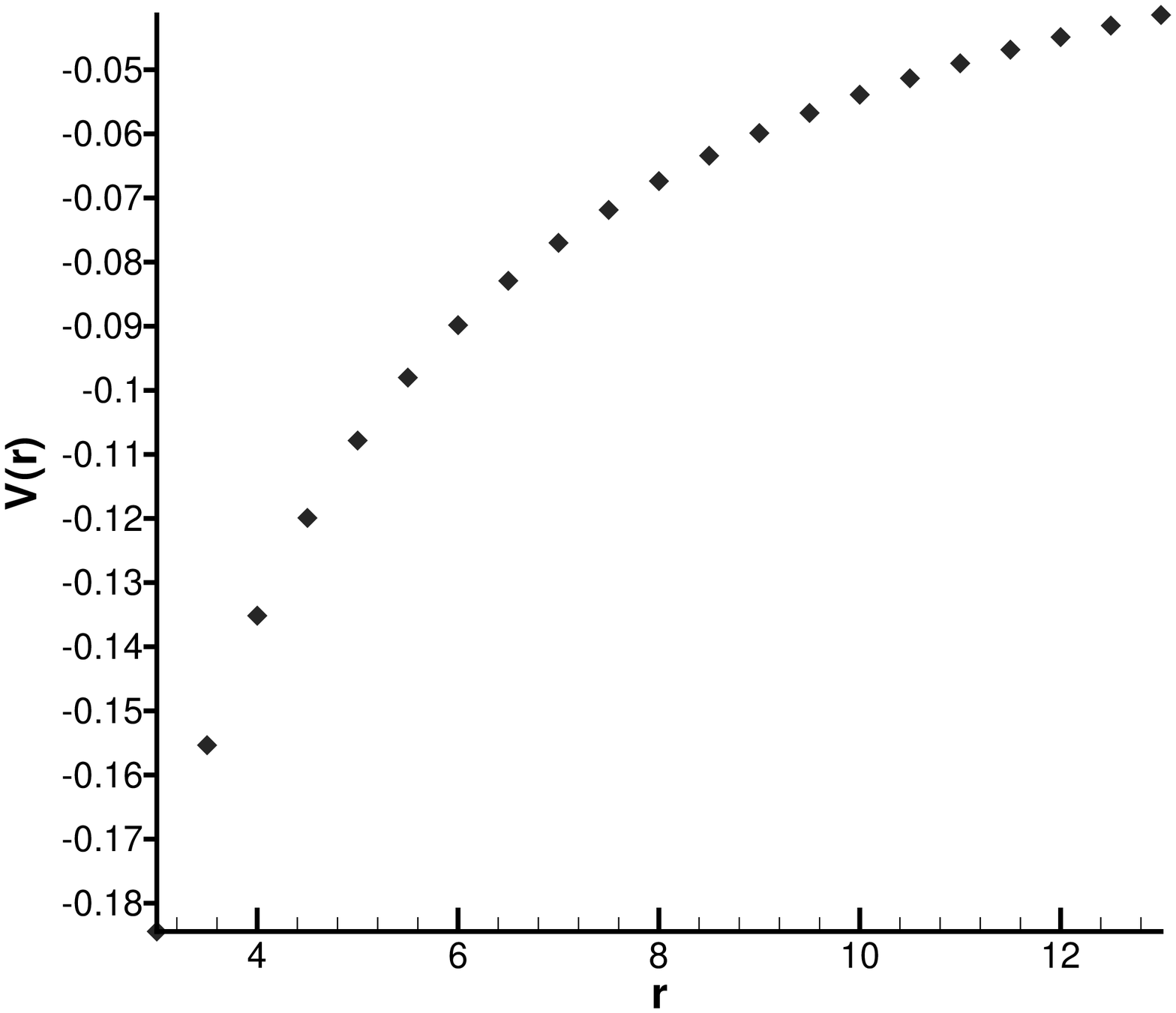,width=10cm}\caption{Brane potential
    for $R_6,\cdots,R_9=\sqrt{\alpha'/2}$, $f=1$ and $p=2$.}\label{ftwo}}

As the branes move closer together,
a condensate of open string tachyons will form, accompanied by
emission of closed string states. We expect the endpoint of this
process to be a stable composite dyonic non-BPS brane, at a
non-trivial minimum of the non-abelian open string tachyon. This kind
of tachyon condensation has been considered before for a pair of
electric non-BPS branes in \cite{lambertsachs}, and from the
supergravity point of view for a large number of coincident branes in
\cite{gian}.

The upshot of the small t expansion is that the 
potential must become attractive
for $r$ near $r_{crit}$.  Combined with the large t expansion
we find that at the very least the potential must
have an unstable equilibrium point for small
enough values of $f$.  

To investigate the potential further we 
performed the integration numerically using 500 digit precision
arithmetic.  
We plot in figures \ref{fone},\ref{ftwo} the
two generic cases that we find for the potential.  Specifically
for the parameters in figure \ref{fone} $f < f_{crit}$, so that the 
potential is asymptotically
repulsive, we find a local maximum in the potential at
some separation of the branes and then an attractive potential
for all smaller separations.  In figure \ref{ftwo} we instead
choose parameters such that $f > f_{crit}$ and find that
the potential is attractive at all separations.  

Finally, let us consider how one might generalize the brane
configuration to realize a potential with a local minimum. 
For purely electric branes, the potential is a monotonic function of
the separation, which is repulsive when the individual branes are
tachyon-free. By
introducing the lower dimensional brane charge we introduce a new length
scale into the interbrane potential proportional to the string length
times a function of the ratio of the charges. As we have seen this is
sufficient to generate a local maximum in the potential. However the
extra charge dominates the behavior at short distances,
leading to a short-range attractive force. By introducing additional
brane charges we introduce additional length scales into the
interbrane potential and in general a local
minimum should be present. 
A challenge for the future is to construct stable non-BPS
brane solutions with these extra charges, which promise new
insights into the brane world scenario.

\acknowledgments

This work was supported in part by DOE
grant DE-FE0291ER40688-Task A.

\end{document}